\documentclass[aps,pra,twocolumn,superscriptaddress,amsmath]{revtex4}
\usepackage{amssymb}
\usepackage{amsthm}
\usepackage{bbm}
\usepackage{enumerate}
\usepackage{color}
\theoremstyle{plain}
\newtheorem{lem}{Lemma}[section] 
\newtheorem{thm}[lem]{Theorem}   
\newtheorem{con}[lem]{Conjecture}   
\theoremstyle{definition}
\newtheorem{defi}{Definition}[section]
\theoremstyle{remark}

\newcommand{\tr}{\operatorname{tr}}
\newcommand{\ket}[1]{\ensuremath{\vert#1\rangle}}

\newcommand{\braket}[2]{\ensuremath{\langle #1\vert#2\rangle}}
\newcommand{\ketbra}[2]{\ensuremath{\vert#1\rangle\!\langle #2\vert}}
\newcommand{\XZ}{\ensuremath{X\!Z}}

\newcommand{\ens}[0]{\ensuremath}
\newcommand{\iE}[0]{\ens{\imath}}
\newcommand{\Eins}[0]{\ens{\mathbbm{1}}}
\newcommand{\F}[0]{\ens{\mathbb{F}}}
\newcommand{\N}[0]{\ens{\mathbb{N}}}
\newcommand{\Z}[0]{\ens{\mathbb{Z}}}
\newcommand{\C}[0]{\ens{\mathbb{C}}}
\newcommand{\Mge}[2]{\ens{\left\lbrace #1|\,#2 \right\rbrace}}
\newcommand{\Mg}[1]{\ens{\left\lbrace #1 \right\rbrace}}
\newcommand{\MgN}[1]{\ens{\Mg{0,\dots,#1}}}
\newcommand{\MgE}[1]{\ens{\Mg{1,\dots,#1}}}
\newcommand{\betrag}[1]{\ens{\left|#1\right|}}

\begin{document}


\title{Complete sets of cyclic mutually unbiased bases in even prime power dimensions}



\author{Oliver Kern}
\affiliation{Institut f\"ur Angewandte Physik, Technische Universit\"at Darmstadt, 64289 Darmstadt, Germany}
\author{Kedar S. Ranade}
\affiliation{Institut f\"ur Angewandte Physik, Technische Universit\"at Darmstadt, 64289 Darmstadt, Germany}
\affiliation{Institut f\"ur Quantenphysik, Universit\"at Ulm, Albert-Einstein-Allee 11, 89069 Ulm, Germany}
\author{Ulrich Seyfarth}
\affiliation{Institut f\"ur Angewandte Physik, Technische Universit\"at Darmstadt, 64289 Darmstadt, Germany}


\date{\today}

\begin{abstract}
We present a construction method for complete sets of cyclic mutually unbiased bases (MUBs) in Hilbert spaces of even prime power dimensions.
In comparison to usual complete sets of MUBs, complete cyclic sets possess the additional property of being generated by a single unitary operator.
The construction method is based on the idea of obtaining a partition of multi-qubit Pauli operators into maximal commuting sets of orthogonal operators with the help of a suitable element of the Clifford group.
As a consequence, we explicitly obtain complete sets of cyclic MUBs generated by a single element of the Clifford group in dimensions $2^m$ for $m=1,2,\dots,24$.
\end{abstract}


\maketitle


\section{Introduction}

One of the basic features of quantum mechanics is that there exist
physical observables which cannot be
measured simultaneously. Given, for example, the measurement outcome of
the $z$-component of the electron
spin, the $x$-component is completely undetermined, i.\,e. given by a
uniform probability distribution.
In mathematical terms, the existence of such measurements arises through
the existence of non-commuting
operators, and one may say that the operators for the $z$- and the
$x$-component are maximally non-commuting,
because measurement of one observable completely destroys the knowledge
of the other. Generalized to
arbitrary finite-dimensional quantum systems, this leads to the concept
of \emph{mutually unbiased bases},
usually abbreviated as MUBs:
Two orthonormal bases of the $d$-dimensional Hilbert space $\mathcal{H} = \mathbb{C}^d$ are said to be \emph{mutually unbiased}, if the absolute value of the inner product of any of the basis vectors of the first basis and any of the basis vectors of the second basis is given by $1/\sqrt{d}$.
MUBs were introduced by Schwinger as ``complementary pairs of
operators'' in order to generate a complete operator basis in a two-state
vector space and an explicit construction method was discussed as early as
1960 \cite{S60}. Twenty years later, Ivanovi\'{c} generalized
this idea to create a complete operator basis of a $d$-dimensional
complex vector space \cite{I81}. The complete measurement (tomography)
of an unknown quantum state motivated his work.
Since a quantum state is described by a density operator that can be
represented by a hermitian matrix with unit trace, the number of real
parameters is $d^2-1$. Every measurement operator can lead to at most
$d$ different outputs, thus there are $d-1$ free parameters due to normalization.
Ivanovi\'{c} consequently stated, that the minimum number of operators
to describe an arbitrary $d$-dimensional quantum state is $d+1$. He gave
an explicit construction method for a ``complete set'' of these
operators in prime dimensions, subsequently. Wootters coined the notion of ``mutual
unbiased bases'' for the different complementary bases \cite{W86}.

As an example for the application of MUBs, consider two observables $A$ and $B$ on $\mathbb{C}^d$
whose eigenbases are mutually unbiased.
Kraus conjectured and Maassen and Uffink have shown, that the optimal uncertainty
relation is given by $H(A) + H(B) \geq \ln d$,
where $H(A)$ and $H(B)$ denote the von Neumann entropies of $A$ and $B$, respectively \cite{K87,MU88}.

For any given dimension $d$ there exists a maximum size for any set consisting of pairwise mutually unbiased bases which is at most $d+1$ \cite{WF89}.
A set of MUBs of this maximum size is called \emph{complete}.
When $d$ is a prime power, i.\,e. $d = q^m$ for a prime $q$ and $m \in \N$, it is known that the maximum size is exactly $d + 1$ and construction methods for complete sets of MUBs are known \cite{WF89,BBRV02,KR03}.
For non-prime-power dimensions the maximum size is unknown, for example in dimension $d=6$ only the lower bound $3$ is known in addition to the upper bound $7$ \cite{BBE06}.

A complete set of cyclic MUBs in dimension $d=2^m$ is a complete set of MUBs, which is fully characterized by a single unitary operator $U$ satisfying $U^{d+1} = \mathbbm{1}_d$, with $\mathbbm{1}_d$ denoting the $d\times d$ identity operator,
such that each of the basis vectors of the different bases is obtained from the basis vectors of the standard basis by the application of powers of $U$.
A complete set of cyclic MUBs can be considered as a generalization of the operator that was used by Gottesman to cyclically transform the three Pauli operators \cite{G98}.
This operator was helpful in security proofs of the six-state protocol \cite{Lo01,GL03}, so we expect that the security of higher dimensional qudit protocols that make use of sets of cyclic MUBs can be proven in a similar way.
Recently, the existence of complete sets of cyclic MUBs in even prime power dimensions was proven by Gow \cite{Gow07} using a representation theoretical argument.
Unfortunately, this method of proof shows merely the existence of such MUBs but does not provide explicit constructions for specific values of $m$.

In this paper, we present such a construction method. It is based on the idea of obtaining a partition of $m$-qubit Pauli operators into maximal commuting sets of orthogonal operators. Starting with a fixed set of commuting operators, the residual sets are generated with the help of a suitable element of the Clifford group. By applying our method, we obtain complete sets of cyclic MUBs in dimensions $2^m$ for $m=1,2,\dots,24$.

In section \ref{sec:Preli}, we start by giving the precise definition of a complete set of cyclic MUBs,
define an equivalence relation for complete sets of MUBs
and introduce the necessary preliminaries for this paper, such as Pauli operators and the Clifford group.
We then describe our construction method in section \ref{sec:constr} and provide complete sets of MUBs for $m \leq 24$ in section \ref{sec:results}.
In section \ref{sec:conclu} we conclude our paper.
In the appendices we provide some analytical results used in the main part.

\section{Basic concepts}\label{sec:Preli}

\subsection{Complete sets of MUBs}

A set of MUBs consists of pairwise mutually unbiased bases, which are defined as follows.
\begin{defi}[Mutually unbiased bases]\hfill\\
  Two orthonormal bases $\mathcal{B}_k=\{ \ket{\psi_0^k}, \dots,
\ket{\psi^k_{d-1}} \}$ and
  $\mathcal{B}_l=\{ \ket{\psi_0^l}, \dots, \ket{\psi^l_{d-1}} \}$
  of the $d$-dimensional Hilbert space $\mathcal{H} = \mathbb{C}^d$ are
said to be \emph{mutually unbiased},
  if there holds
  \begin{equation*}
    \vert \braket{\psi_i^k}{\psi_j^l} \vert = 1/\sqrt{d}
  \end{equation*}
  for all $0\leq i,j<d$.
\end{defi}

\noindent
For any given dimension $d$ there exists a maximum size $N(d)$ for any set of MUBs, and it is known that $N(d) \leq d+1$ \cite{WF89}.
\begin{defi}[Complete sets of MUBs]\hfill\\
In the $d$-dimensional Hilbert space $\mathcal{H} = \mathbb{C}^d$, a set of MUBs of the maximum size $N(d)$ is called \emph{complete}. 
\end{defi}

\noindent
When $d$ is a prime power, i.\,e. $d = p^m$ for a prime $p$ and $m \in \N$, it is known that $N(d) = d + 1$ and construction methods for complete sets of MUBs are known \cite{WF89,BBRV02,KR03}, while for non-prime-power dimensions $N(d)$ is unknown.

\begin{defi}[Complete sets of cyclic MUBs]\hfill\\
In dimension $d=2^m$ \emph{a complete set of cyclic MUBs} is a \emph{complete set of MUBs} $\{\mathcal{B}_1,\dots,\mathcal{B}_{d+1} \}$, with
$\mathcal{B}_1=\{ \ket{0}, \ket{1}, \dots, \ket{d-1} \}$
denoting the standard basis \footnote{To be precise, a complete set of cyclic MUBs does not need to contain the standard basis, but it can always be brought into such a form by a unitary transformation.}, which is fully characterized by a single unitary operator $U$ satisfying $U^{d+1} = \mathbbm{1}_d$, with $\mathbbm{1}_d$ denoting the $d\times d$ identity operator,
as follows:
Each of the basis vectors of the bases
$\mathcal{B}_k=\{ \ket{\psi_0^k}, \dots, \ket{\psi^k_{d-1}} \}$ (with $2\leq k \leq d+1$) is obtained from the basis vectors of $\mathcal{B}_1$ by the application of powers of $U$ such that $\ket{\psi_i^k} = U^{k-1} \ket{i}$.
\end{defi}

\subsection{Equivalence of MUBs}\label{subsec:uniequi}
Let us assume that we have two complete sets of MUBs $\{\mathcal{B}_1,\dots,\mathcal{B}_{d+1} \}$ and $\{\mathcal{A}_1,\dots,\mathcal{A}_{d+1} \}$ in a Hilbert space $\mathcal{H} = \mathbb{C}^d$ of prime power dimension $d=q^m$.
We will employ the convention to write the components $b_{i,j}^k$ of the basis vectors
\begin{equation}
 \ket{\psi_i^k} = \sum_{j=0}^{d-1} b_{i,j}^k \ket{ j }
\end{equation}
of a basis $\mathcal{B}_k = \{ \ket{\psi_0^k},\dots, \ket{\psi_{d-1}^k} \}$ in the columns of a matrix $B_k$,
\begin{equation}
B_k=\begin{pmatrix}
 b_{0,0}^k    & b_{1,0}^k   & \hdots & b_{d-1,0}^k \\
 b_{0,1}^k    & b_{1,1}^k   &        & b_{d-1,1}^k \\
 \vdots       &             &        & \vdots      \\
 b_{0,d-1}^k  & b_{1,d-1}^k & \hdots & b_{d-1,d-1}^k
\end{pmatrix},
\end{equation}
i.\,e. for the standard basis $\mathcal{B}_1$, we obtain $B_1=\mathbbm{1}_d$.
The two sets of MUBs are equivalent if there exists some fixed unitary $V$ taking one set into the other.
There are two caveats:
First, any of the basis vectors is fixed only up to an arbitrary global phase. Second, the order of the basis vectors of a certain basis is irrelevant as is the order of the different bases.
Hence, we have the following definition:
\begin{defi}[Equivalence of MUBs]\label{def:uequi}\hfill\\
Two complete sets of MUBs $\{\mathcal{B}_1,\dots,\mathcal{B}_{d+1} \}$ and $\{\mathcal{A}_1,\dots,\mathcal{A}_{d+1} \}$ are said to be equivalent, if there exists a unitary matrix $V \in M_d(\C)$, matrices $W_k$, $k \in \MgE{d+1}$, which contain exactly one non-zero entry per row and column, the absolute value of which must be unity, and a permutation $\pi$ on $\MgE{d+1}$, such that there holds
\begin{equation*}
  A_k = V B_{\pi(k)} W_k
\end{equation*}
for all values $k \in \MgE{d+1}$.
\end{defi}

\subsection{Pauli operators}

We start with the definition of Pauli operators acting on a one-qudit Hilbert space $\mathcal{H}_q=\mathbb{C}^q$ of prime dimension $q$.
The Pauli $X$ and $Z$ operators are defined by
\begin{align}
 X\ket{i} &= \ket{i+1\!\!\!\pmod{q}}\\
 Z\ket{i} &= \omega^i \ket{i},
\end{align}
where $\omega = \exp(2\pi \iE/q)$ denotes a complex primitive $q$-th root of unity.
It follows that $ZX=\omega XZ$.
For any vector
$\vec{a} = (\vec{a}^x\vert\vec{a}^z) = (a^x_1,\dots,a^x_m\vert a^z_1,\dots,a^z_m) \in \mathbb{F}_q^{2m}$, let the Pauli operator $\XZ(\vec{a})$ acting on the $m$-qudit Hilbert space $\mathcal{H} = \mathcal{H}_q^{\otimes m}$ of dimension $d=q^m$ be defined by
\begin{equation}
 \XZ(\vec{a}) = \begin{cases}
     \iE^{a^x_1a^z_1}X^{a^x_1}Z^{a^z_1}\otimes\dots\otimes \iE^{a^x_ma^z_m}X^{a^x_m}Z^{a^z_m}&, q=2\\
     X^{a^x_1}Z^{a^z_1}\otimes\dots\otimes X^{a^x_m}Z^{a^z_m}&, q\geq 3
               \end{cases},
\end{equation}
so that the eigenvalues of $\XZ(\vec{a})$ are powers of $\omega$.
%
%
If we represent the one-qubit Pauli operators in the standard basis, we obtain the well known Pauli matrices,
\begin{align}
 \XZ(0\vert 0) &= \begin{pmatrix} 1&0\\ 0&1 \end{pmatrix} &
 \XZ(1\vert 0) &= \begin{pmatrix} 0&1\\ 1&0 \end{pmatrix} \\
 \XZ(1\vert 1) &= \begin{pmatrix} 0&-\iE\\ \iE&0 \end{pmatrix} &
 \XZ(0\vert 1) &= \begin{pmatrix} 1&0\\ 0&-1 \end{pmatrix},
\end{align}
which we will also denote as $\mathbbm{1}_2,X,Y$ and $Z$.
%
%
For $q\geq 3$ we obtain
\begin{equation}
 \XZ(\vec{a}) \cdot \XZ(\vec{b}) = \omega^{\sum_i a^z_ib^x_i } \XZ(\vec{a}+\vec{b}),
\end{equation}
while for $q=2$ this expression holds up to powers of $\iE$.
As a consequence, $\XZ(\cdot)$ gives rise to a unitary projective representation of $\mathbb{F}_q^{2m}$,
which by itself forms a group under addition modulo $q$.
We denote the set containing all $m$-fold tensor products of Pauli operators as
\begin{equation}
 \mathcal{P}_q^m = \{ \XZ(\vec{a}) \ \vert\  \vec{a}\in\mathbb{F}_q^{2m} \}.
\end{equation}
Finally, the symplectic inner product between elements $\vec{a}$ and $\vec{b}$ of $\mathbb{F}_q^{2m}$ is defined as
\begin{equation}
(\vec{a},\vec{b})_{sp} = \sum_{i=1}^m a^z_i b^x_i - a^x_i b^z_i \pmod{q}.
\end{equation}
With the help of the inner product defined above, the order of a product of two Pauli operators $\XZ(\vec{a})$ and $\XZ(\vec{b})$ can be inverted,
\begin{equation}
 \XZ(\vec{a}) \cdot \XZ(\vec{b})  = \omega^{(\vec{a},\vec{b})_{sp}}  \XZ(\vec{b}) \cdot \XZ(\vec{a}).
\end{equation}
It follows that two Pauli operators $\XZ(\vec{a})$ and $\XZ(\vec{b})$ commute if and only if the symplectic inner product between $\vec{a}$ and $\vec{b}$ vanishes.

\subsection{Clifford group operators}\label{subsec:cliff}

We consider a $d=q^m$ dimensional Hilbert space $\mathcal{H}$ of $m$ qudits of dimension $q$.
The Clifford group $\mathfrak{C}^m_q$ on $\mathcal{H}$ is defined as the group of unitary operators $U$ which map $m$-qudit Pauli operators onto $m$-qudit Pauli operators \cite{G98-2},
\begin{multline}
\mathfrak{C}^m_q = \{U \in M_d(\mathbb{C}) \text{ unitary} \ \vert\ \\
(\forall \vec{a} \in \F_q^{2m})(\exists \vec{a}^\prime \in \F_q^{2m})(U \XZ(\vec{a}) U^\dagger = \XZ(\vec{a}^\prime) \},
\end{multline}
with $M_d(\mathbb{C})$ denoting the set of $d\times d$ matrices with entries in $\mathbb{C}$.

Any member $U$ of the Clifford group is fully specified when the action
$\XZ(\vec{a}') = U \XZ(\vec{a}) U^\dagger$ of $U$ on a generating set of elements of the Pauli group $\mathcal{P}_q^m$ is known.
In the following we assume that such a generating set is given by the operators $\XZ(\vec{x}_i)$ and $\XZ(\vec{z}_i)$ with
$\vec{x}_i = (0,\dots,0,1,0,\dots 0 \vert 0,\dots,0 )\in \mathbb{F}_q^{2m}$ having a one in position $1\leq i\leq m$ and
$\vec{z}_i = (0,\dots,0 \vert 0,\dots,0,1,0,\dots 0 )\in \mathbb{F}_q^{2m}$ having a one in position $m+1\leq m+i\leq 2m$.
To understand this fact, we note that any $\vec{a}\in\mathbb{F}_q^{2m}$ can be expressed as a linear superposition of these generators,
$\vec{a} = \sum_i a^x_i \vec{x}_i + \sum_i a^z_i \vec{z}_i$.
As a consequence, we obtain
\begin{align}
U \XZ(\vec{a}) U^\dagger &= \prod_{i=1}^m ( U\XZ(\vec{x}_i) U^\dagger)^{a^x_i}
\prod_{i=1}^m ( U\XZ(\vec{z}_i) U^\dagger)^{a^z_i} \nonumber\\
&=  \XZ(\vec{a}') 
\end{align}
with $\vec{a}' = \sum_i a^x_i \vec{x}'_i + \sum_i a^z_i \vec{z}'_i$.
It follows that the mapping of a generator $\vec{g}\in\{ \vec{x}_1,\dots,\vec{x}_m,\vec{z}_1,\dots,\vec{z}_m \}$ onto its image $\vec{g}'$ can be described as $\vec{g}'^T = C \cdot \vec{g}^T$ using a $2m\times 2m$ matrix $C\in M_{2m}(\mathbb{F}_q)$ whose first $m$ columns contain the transposed row vectors $\vec{x}'_i$ and whose second $m$ columns contain the transposed $\vec{z}'_i$.
In addition, the image $\vec{a}'$ of an arbitrary element $\vec{a}\in\mathbb{F}_q^{2m}$ can easily be expressed as $\vec{a}'^T= C\cdot \vec{a}^T$.
Since the commutator relations for the $\XZ(\vec{x}_i)$ and $\XZ(\vec{z}_i)$, namely
\begin{align}
 ( \vec{x}_i, \vec{x}_j )_{sp} &= 0, &
 ( \vec{z}_i, \vec{z}_j )_{sp} &= 0, &
 ( \vec{z}_i, \vec{x}_j )_{sp} &= \delta_{ij},
\end{align}
for all $1\leq i,j\leq m$, have to remain unchanged for the $\XZ(\vec{x}'_i)$ and $\XZ(\vec{z}'_i)$,
the matrix $C$ underlies the constraint
\begin{equation}\label{eq:Csympl}
C^T \cdot \begin{pmatrix}0_m&-\mathbbm{1}_m\\\mathbbm{1}_m&0_m\end{pmatrix} \cdot C = \begin{pmatrix}0_m&-\mathbbm{1}_m\\\mathbbm{1}_m&0_m\end{pmatrix} \pmod q,
\end{equation}
and is called symplectic.

If only the matrix $C$ describing the action of a Clifford unitary $U$ is known, the matrix $U$ can be reconstructed as follows:
The first $m$ columns of $C$ contain the transposed of the images $\vec{x}'_i$ of the generators $\vec{x}_i$, while columns $m+1,\dots,2m$ contain the transposed of the images $\vec{z}'_i$ of the generators $\vec{z}_i$ (with $1\leq i \leq m$).
Let us use the corresponding commuting Pauli operators $\XZ(\vec{z}'_i)$ to define the so-called stabilizer state $\ket{ \vec{0} }_L$ as the common eigenvector of eigenvalue $+1$,
\begin{equation}
  \XZ(\vec{z}'_i) \ket{ \vec{0} }_L = +1 \cdot \ket{ \vec{0} }_L, \text{ for all } 1\leq i \leq m.
\end{equation}
We obtain a so-called logical orthonormal basis $\mathcal{B}_L =\{ \ket{\vec{0}}_L, \ket{\vec{1}}_L, \dots, \ket{\overrightarrow{2^m-1}}_L \}$, with $\vec{i}=(i_1,\dots,i_m)\in\mathbb{F}_2^m$ and $i=\sum_{j=1}^m i_j\cdot 2^{m-j}$, by applying the operator $\prod_j \XZ( i_j \cdot \vec{x}'_j )$ onto $\ket{\vec{0}}_L$.
It is easy to verify that
\begin{align}
 \XZ(\vec{z}'_j)  \ket{\vec{i}}_L &= \omega^{i_j}  \ket{\vec{i}}_L \\
 \XZ(\vec{x}'_j)  \ket{\vec{i}}_L &= \ket{ (i_1,\dots,i_j+1,\dots,i_m) }_L,
\end{align}
which is why the $\XZ(\vec{z}'_j)$ and $\XZ(\vec{x}'_j)$ are called logical Pauli $Z_j$ and $X_j$ operators, respectively.
Expressed in the standard basis, the $d\times d$ unitary matrix $U = (c_{jk} )$ contains in its $k$-th column the components of the vector $\ket{\vec{k}}_L = \sum_j c_{jk} \ket{\vec{j}}$ and satisfies $U\ket{ \vec{k} } = \ket{ \vec{k} }_L$.

\section{Construction method}\label{sec:constr}

\subsection{Complete MUBs and maximal commuting operator bases}

Let $M_d(\mathbb{C})$ denote the set of all $d\times d$ matrices with entries in $\mathbb{C}$.
Two matrices $A$ and $B$ from the set $M_d(\mathbb{C})$ are said to be orthogonal if their trace inner product $\langle A, B\rangle = \tr(A^\dagger B)$ vanishes.
A maximal commuting unitary operator basis for $M_d(\mathbb{C})$ is a set $\mathcal{M}=\{ u_1,\dots,u_{d^2} \}$ of unitary matrices containing the identity matrix $\mathbbm{1}_d$
that can be partitioned as $\mathcal{M}=\Mg{\Eins_d}\cup \mathcal{C}_1\cup\dots\cup\mathcal{C}_{d+1}$ into $d+1$ disjoint sets $\mathcal{C}_j$ containing $d-1$ commuting operators each.
The following theorem due to Bandyopadhyay et~al. \cite{BBRV02} allows the construction of a complete set of MUBs in prime power dimensions with the help of a maximal commuting unitary operator basis that consists of pairwise orthogonal operators only.

\begin{thm}[Construction of MUBs]\hfill\\
A maximal commuting unitary operator basis for $M_d(\mathbb{C})$ consisting of pairwise orthogonal operators
defines a complete set of $d+1$ MUBs.
\end{thm}
\begin{proof}
Each of the sets $\mathcal{C}_j'=\Mg{\Eins_d}\cup\mathcal{C}_j$ contains $d$ orthogonal and commuting unitary operators which define a common eigenbasis $\mathcal{B}_j=\{ \ket{\psi_1^j},\dots,\ket{\psi_d^j} \}$ (unique up to phases).
Let us denote the elements of $\mathcal{C}_j'$ as $\mathcal{C}_j'=\{ u_{j,0},u_{j,1},\dots, u_{j,d-1} \}$ with $u_{j,0}=\mathbbm{1}_d$. Expressing these elements in terms of the eigenbasis $\mathcal{B}_j$ leads to the diagonal representations $u_{j,t} = \sum_{k=1}^d \lambda_{j,t,k} \ketbra{\psi_k^j}{\psi_k^j}$ ($0\leq t \leq d-1$).
Using the orthogonality of the unitaries $u_{j,t}$, we obtain the equation
\begin{equation*}
d \delta_{t,0}\delta_{t',0} = \tr( u_{j,t}^\dagger \cdot u_{j',t'} ) = \sum_{k,k'=1}^d \lambda^\ast_{j,t,k} \lambda_{j',t',k'} \vert\braket{\psi_k^j}{\psi_{k'}^{j'}}\vert^2
\end{equation*}
for $0\leq t,t' \leq d-1$ and $1\leq j < j' \leq d+1$.
Defining the unitary $d\times d$ matrices $M_j=(m_{tk})$ with entries $m_{tk}=\lambda_{j,t,k}/\sqrt{d}$, the above equation can be written as $M_j^\ast\otimes M_{j'} \cdot \vec{v}^T = (1,0,\dots,0)^T$ with vector
$\vec{v}=(
\vert\braket{\psi_1^j}{\psi_1^{j'}}\vert^2,
\vert\braket{\psi_1^j}{\psi_2^{j'}}\vert^2,
\dots,
\vert\braket{\psi_d^j}{\psi_d^{j'}}\vert^2 )$.
Inversion of this vector equation leads to $\vec{v} = (1/d,1/d,\dots,1/d)$ and hence shows that the bases $\mathcal{B}_j$ and $\mathcal{B}_{j'}$ are mutually unbiased.
Since this proof applies to all $1\leq j < j' \leq d+1$, the set of eigenbases $\{ \mathcal{B}_1,\dots, \mathcal{B}_{d+1} \}$ forms a complete set of MUBs.
\end{proof}

\subsection{Pauli operators and maximal commuting operator bases}

As it is discussed in \cite{BBRV02}, in prime power dimensions $d=q^m$ the set $\mathcal{P}_q^m$ of Pauli operators can always be partitioned as $\mathcal{P}_q^m = \Mg{\Eins_d}\cup \mathcal{C}_1\cup\dots\cup\mathcal{C}_{d+1}$ in order to form a maximal commuting operator basis for $d\times d$ matrices.
Let us give an explicit example for $q=2$ and $m=2$:
\begin{equation}\label{eq:exd4}
\begin{array}{cc@{}c@{\otimes}c@{\ ,\ }c@{\otimes}c@{\ ,\ }c@{\otimes}c@{\ ,}r@{\otimes}c@{}c}
\Mg{\Eins_4}\cup\mathcal{C}_1 &=\{& \mathbbm{1}_2&\mathbbm{1}_2 & Z&\mathbbm{1}_2 & \mathbbm{1}_2&Z & Z&Z & \} \\
\Mg{\Eins_4}\cup\mathcal{C}_2 &=\{& \mathbbm{1}_2&\mathbbm{1}_2 & X&\mathbbm{1}_2 & \mathbbm{1}_2&X & X&X & \} \\
\Mg{\Eins_4}\cup\mathcal{C}_3 &=\{& \mathbbm{1}_2&\mathbbm{1}_2 & Y&X & X&Z & -Z&Y & \} \\
\Mg{\Eins_4}\cup\mathcal{C}_4 &=\{& \mathbbm{1}_2&\mathbbm{1}_2 & Y&Y & Y&\mathbbm{1}_2  & \mathbbm{1}_2&Y & \} \\
\Mg{\Eins_4}\cup\mathcal{C}_5 &=\{& \mathbbm{1}_2&\mathbbm{1}_2 & Y&Z & Z&X & -X&Y & \}
\end{array}
\end{equation}
The common eigenbasis of the operators in $\Mg{\Eins_4}\cup\mathcal{C}_1$ is the standard basis $\mathcal{B}_1=\{ \ket{00},\ket{01},\ket{10},\ket{11} \}$ of $m=2$ qubits, or equivalently the standard basis $\mathcal{B}_1=\{ \ket{0},\ket{1},\ket{2},\ket{3} \}$ of one qudit of dimension $d=2^m$.
The common eigenbasis $\mathcal{B}_2$ of $\Mg{\Eins_4}\cup\mathcal{C}_2$ consists of the basis vectors
$\ket{\psi^2_i} = \sum_{j=0}^3 b^2_{i,j} \ket{j}$.
Let us write the components of the $\ket{\psi^2_i}$ in the columns of a matrix $B_2 = (b^2_{k,j})$, and we obtain
\begin{equation}
 B_2 = \frac{1}{2}
\begin{pmatrix}
\iE & \iE & 1 &-1\\
\iE &-\iE & 1 & 1\\
\iE & \iE &-1 & 1\\
\iE &-\iE &-1 &-1
\end{pmatrix},
\end{equation}
where we used a special choice of global phases for the $\ket{\psi^2_i}$.
Setting $U=B_2$, it can be verified that the remaining bases $\mathcal{B}_3$, $\mathcal{B}_4$ and $\mathcal{B}_5$ are given by the matrices $B_3=U^2$, $B_4=U^3$ and $B_5=U^4$ (also note that $U^5=\mathbbm{1}_4$).
This means that our example describes a complete set of cyclic MUBs in dimension $d=4$.
Since we are interested in the construction of cyclic MUBs for $q=2$, the question is how such a partition of $\mathcal{P}_2^m$ can be obtained for arbitrary $m$.
Unfortunately, the method for the construction of partitions of $\mathcal{P}_q^m$ mentioned in \cite{BBRV02} does not lead to cyclic MUBs in general.

\subsection{Construction of cyclic MUBs}\label{subsec:constr}

As it can be seen from the example for $m=2$ given in equation \eqref{eq:exd4}, a complete set of cyclic MUBs in dimension $d=2^m$ can be obtained if we find a partition of the Pauli operators $\mathcal{P}_2^m$ into $d+1$ disjoint sets $\mathcal{C}_j$ of size $d-1$ containing commuting operators, such that
\begin{equation}\label{eq:Cjp1_UCjU}
 \mathcal{C}_{j+1} = U \mathcal{C}_j U^\dagger
\end{equation}
for some unitary $U$ with $U^{d+1}=\mathbbm{1}_d$.
In this case the common eigenbases $\mathcal{B}_j$ of the operators $\mathcal{C}_j$ are simply obtained by applying the unitaries $U^{j-1}$ onto the elements of the common eigenbasis $\mathcal{B}_1$ of the operators $\mathcal{C}_1$.
Let us assume now that we always choose $\mathcal{C}_1$ to consist of all $d-1$ tensor products of $Z$ operators, i.\,e. of all Pauli operators $\XZ(\vec{a})$ such that $\vec{a}=(\vec{0}\vert \vec{a}^z )$ with $\vec{a}^z\in \mathbb{F}_2^m\setminus\{ \vec{0} \}$ and $\vec{0}=(0,\dots,0)$.
In this case the common eigenbasis of $\Mg{\Eins_d}\cup\mathcal{C}_1$ is always the standard basis $\mathcal{B}_1 = \{ \ket{0\dots 00}\equiv\ket{0}, \ket{0\dots 01}\equiv\ket{1}, \dots, \ket{1\dots 11}\equiv\ket{d-1} \}$.
As in subsection \ref{subsec:uniequi}, we will employ the convention to write the components of the basis vectors of a basis
$\mathcal{B}_k = \{ \ket{\psi_0^k},\dots, \ket{\psi_{d-1}^k} \}$ in the columns of a matrix $B_k$.

Note that we can store the exponentially many members of $\mathcal{C}_1$ in an efficient form by writing $\Mg{\Eins_d}\cup\mathcal{C}_1 = \{ \XZ(\vec{a}) \vert \vec{a} = \vec{c}\cdot C_1 \text{ with } \vec{c}\in\mathbb{F}_2^m  \}$ with the $m\times 2m$ generator matrix $C_1 = ( 0_m \vert \mathbbm{1}_m )$.
A unitary $U$ which generates the remaining sets $\mathcal{C}_j$ with $j\geq 2$ via \eqref{eq:Cjp1_UCjU} maps $m$-qubit Pauli operators onto $m$-qubit Pauli operators and hence is a member of the Clifford group.

Using the representation of a Clifford group unitary $U$ in terms of a symplectic matrix $C\in M_{2m}(\mathbb{F}_2)$, we can reformulate condition \eqref{eq:Cjp1_UCjU} as follows:
Let the set $\Mg{\Eins_d}\cup\mathcal{C}_j$ be specified by a $m\times 2m$ generator matrix $C_j$,
i.\,e. $\Mg{\Eins_d}\cup\mathcal{C}_j = \{ \XZ(\vec{a}) \vert \vec{a} = \vec{c}\cdot C_j \text{ with } \vec{c}\in\mathbb{F}_2^m  \}$, then the set $\mathcal{C}_{j+1}$
has to be specified by the generator matrix $C_{j+1}=  C_j \cdot C^T$.

Let us summarize the results so far.
We are interested in finding a partition of the set of Pauli operators $\mathcal{P}_2^m = \Mg{\Eins_d}\cup\mathcal{C}_1\cup\dots\cup\mathcal{C}_{d+1}$ into disjoint sets $\mathcal{C}_j$ of size $d-1$ containing commuting operators, such that $\mathcal{C}_{j+1} = U \mathcal{C}_j U^\dagger$ for a Clifford unitary $U$ with $U^{d+1}=\mathbbm{1}_d$.
Fixing the first set $\mathcal{C}_1$ by choosing the generator matrix $C_1=(0_m\vert \mathbbm{1}_m)$ leads to the basis $B_1=\mathbbm{1}_d$, and the remaining bases are specified by the matrices $B_j=U^{j-1}$, or in other words by the basis $U=B_2$. If in addition $U^{d+1}=\mathbbm{1}_d$, we have a complete set of cyclic MUBs in dimension $d=2^m$ which is specified by the powers $\{ U^j \vert  0\leq j\leq d\}$ of a single matrix $U$.
Instead of looking for such a $d\times d$ dimensional Clifford unitary $U$ directly, it is easier to look for its $2m\times 2m$ dimensional representation $C\in M_{2m}(\mathbb{F}_2)$:
We have to find a $C\in M_{2m}(\mathbb{F}_2)$ such that
\begin{enumerate}[I.)]
 \item $C$ satisfies equation \eqref{eq:Csympl} (i.\,e. $C$ is symplectic),
 \item the generator matrices $C_j = C_1 \cdot (C^{j-1})^T$ (with $1\leq j \leq d+1$ and $C_1=(0_m\vert \mathbbm{1}_m)$) span non-overlapping vector spaces,
 \item $C_{d+2} = C_1$, or in other words $C^{d+1}=\mathbbm{1}_{2m}$.
\end{enumerate}
If such a $C$ is found, the last step is to construct the unitary $U$ corresponding to $C$.

\subsection{Finding a suitable $C$}\label{subsec:findc}

Let us now describe how for every $m\in \mathbb{N}$ a $2m\times 2m$ matrix $C$ with entries in $\mathbb{F}_2$ satisfying the three conditions I.), II.), and III.) stated at the end of the last subsection can be found.
We start with the assumption that there always exists such a matrix $C$ having the form
\begin{equation}\label{eq:CgleichB110}
 C = \begin{pmatrix}
       B & \mathbbm{1}_m \\
       \mathbbm{1}_m & 0_m
     \end{pmatrix},
\end{equation}
with $B \in M_m(\mathbb{F}_2)$.
Note that in order for $C$ to satisfy condition I.) (being symplectic), $B$ has to be symmetric, i.\,e. $B=B^T$.
Equation \eqref{eq:CgleichB110} allows us to write $C^n$ as
\begin{equation}\label{eq:Cn-fk}
 C^n = \begin{pmatrix}
       f_n(B) & f_{n-1}(B) \\
       f_{n-1}(B) & f_{n-2}(B)
     \end{pmatrix},
\end{equation}
using the recursively defined polynomials $f_n$ over the finite field $\mathbb{F}_2$ satisfying
\begin{equation}
  f_n(x) = f_{n-1}(x)\cdot x + f_{n-2}(x),
\end{equation}
with $f_{-1}(x)=0$, $f_0 (x)=1$ and $f_1(x)=x$.
We prove some properties of the $f_n$ we are going to use in the following in appendix \ref{sec:propfn}.

Starting with the generator matrix $C_1 = (0_m\vert \mathbbm{1}_m)$ generating the set $\Mg{\Eins_d}\cup\mathcal{C}_1$, our particular choice of $C$ leads to the generator list $\{ C_1, C_2, \dots, C_{d+1} \}$ with $C_j = C_1 \cdot (C^{j-1})^T = ( f_{j-2}(B) \vert f_{j-3}(B) )$.
Now in order to satisfy condition II.), any two generators $C_j$ and $C_k$ (with $1\leq j < k \leq d+1$ and $d=2^m$) have to span non-overlapping vector spaces, or in other words the $2m\times 2m$ matrix
\begin{equation}\label{eq:cjck}
\begin{pmatrix}
  f_{j-2}(B) & f_{j-3}(B) \\
  f_{k-2}(B) & f_{k-3}(B)
\end{pmatrix}
\end{equation}
with entries in $\mathbb{F}_2$ has to be invertible; this can be checked with the help of lemmas \ref{lem:fjk} and \ref{lem:invmatrices} of the appendix as follows:
According to lemma \ref{lem:invmatrices} a $2m\times 2m$ block matrix as \eqref{eq:cjck} is invertible if and only if the $m\times m$ matrix
$f_{j-2}(B) \cdot  f_{k-3}(B) - f_{j-3}(B)\cdot f_{k-2}(B) = f_{\vert j-k \vert -1}(B) \in M_m(\mathbb{F}_2)$ is invertible, where the latter identity is obtained by lemma \ref{lem:fjk}.
Hence, all $f_{j}(B)$ with $1\leq j \leq   d-1$ have to be invertible.
Finally, in order to satisfy condition III.), we have to demand that $C_{d+2} = ( f_{d}(B) \vert f_{d-1}(B) ) = C_1$ with $d=2^m$.
This latter condition demands (due to the fact that $f_{n-2}(x) = f_{n-1}(x)\cdot x + f_n(x)$) that the equation $f_{d-1-j}(B)=f_j(B)$ holds for any $j=1,\dots, 2^{m-1} $. It is interesting to note that the polynomials $f_{d-1}(B)$ and $f_{d-2}(B)$ have simple forms, see lemmata \ref{lem:f2m1} and \ref{lem:f2m2}, e.g. $f_{d-1}(B)=B^{d-1}$. Combined with the previous conditions, we obtain the following conditions which are faster to verify:
\begin{enumerate}[i.)]
 \item $B=B^T$,
 \item $f_{j}(B)$ is invertible for all $1\leq j \leq 2^{m-1}$,
 \item $f_{2^{m-1}}(B)=f_{2^{m-1}-1}(B)$.
\end{enumerate}

\subsubsection{Construction of $U$}
So far, we showed that in order to find a symplectic $C$ for a fixed value of $m\in \mathbb{N}$, it suffices to find a $B\in M_m(\mathbb{F}_2)$ satisfying the above conditions.
Before we proceed to explain how we found such matrices $B$ for different values of $m$ up to $m=24$, let us construct the unitary Clifford operator $U$ corresponding to a matrix $C$ of the form \eqref{eq:CgleichB110} in the way it was explained in subsection \ref{subsec:cliff}.
The stabilizer state $\ket{\vec{0}}_L$ is defined as the common eigenvector with eigenvalue $+1$ of the logical Pauli $Z_j$ operators, which are now simply given by the usual Pauli $X_j$ operators.
Hence,
\begin{equation}
 \ket{\ \vec{0}\ }_L = \frac{1}{\sqrt{2^m}} \sum_{ \mathbb{F}_2^m\ni\vec{i} = (0,\dots,0) }^{(1,\dots,1)} \ket{\ \vec{i}\ }.
\end{equation}
To obtain the $\ket{\ \vec{j}\ }_L \equiv \ket{(j_1,\dots,j_m)}_L$, we have to apply the operator $\prod_k \XZ(\vec{x}'_k)^{j_k}$, where the $\vec{x}_k'$ are given by
\begin{equation}
 \vec{x}_k' = ( B_{k1},\dots,B_{km} \vert \delta_{1k},\dots,\delta_{mk} ) \in \mathbb{F}_2^{2m},
\end{equation}
and it follows that
\begin{equation}
 \ket{\ \vec{j}\ }_L = \frac{1}{\sqrt{2^m}} \sum_{ \mathbb{F}_2^m\ni\vec{i} = (0,\dots,0) }^{(1,\dots,1)} p_{\vec{j}} (-1)^{\vec{i}\cdot\vec{j}} \ket{\ \vec{i}\ },
\end{equation}
where the phases $p_{\vec{j}}\in\{\pm 1,\pm \iE\}$ are obtained from $B$ as follows:
Let $\vec{b}=(B_{11},\dots,B_{mm}) \in \mathbb{F}_2^m$ be the diagonal of $B$, let $\vec{B}_k=(B_{k1},\dots,B_{km})\in \mathbb{F}_2^m$ be the $k$-th row of $B$, and let $\vec{v}_{\rightarrow k} = ( v_1,\dots,v_k,0,\dots,0)$ for any vector $\vec{v}=(v_1,\dots,v_k,v_{k+1},\dots,v_m)$.
Then, (with $\exp_x(y) = x^y$)
\begin{equation}\label{eq:pj}
 p_{\vec{j}} = \exp_\iE\bigl( \vec{b}\cdot\vec{j} \bigr) \cdot \exp_{-1}\Bigl( \sum_{k=1}^m j_k\cdot ( \vec{B}_k\cdot\vec{j}_{\rightarrow k} ) \Bigr).
\end{equation}
Since $U$ contains the components of the $ \ket{\ \vec{j}\ }_L$ as columns, we can write it as
\begin{equation}\label{eq:UHdiap}
 U = H^{\otimes m} \cdot \operatorname{diag}( p_{\vec{0}}, p_{\vec{1}}, \dots, p_{\overrightarrow{2^m-1}} ) \bigr) \cdot e^{\iE\psi},
\end{equation}
where $H^{\otimes m}$ denotes the $m$-fold tensor product of the Hadamard matrix
\begin{equation}
 H=\frac{1}{\sqrt{2}}\begin{pmatrix}
  1 & 1 \\
  1 & -1
 \end{pmatrix}.
\end{equation}
If we assume that our conjecture \ref{SpectrumConjecture} is valid, we can choose the trace of our cyclic $U$
of order $2^m+1$ to be equal to $-1$ and apply a global phase $e^{\iE\psi}$, which is determined by
\begin{equation}\label{eq:globalbytr}
   e^{\iE\psi}= -\tr \bigl( H^{\otimes m} \cdot \operatorname{diag}( p_{\vec{0}}^\ast, p_{\vec{1}}^\ast, \dots, p_{\overrightarrow{2^m-1}}^\ast ) \bigr),
\end{equation}
where $p_{\vec{j}}^\ast$ denotes the complex conjugate of $p_{\vec{j}}$.

\subsubsection{Search for $B$}
Even though the number $2^{m(m+1)/2}$ of symmetric matrices $B\in M_m(\mathbb{F}_2)$ seems to be rather large for a complete search and large $m$, it turns out that a suitable $B$ can quite easily be found for moderate values of $m$.
%
For $m\geq 4$, we make the guess that a suitable matrix $B=(b_{ij})$ exists with entries $b_{ij} = \beta_{ij} + \alpha_{ij}$ for $1\leq i,j\leq m$, with
\begin{equation}
 \beta_{ij}= \begin{cases} 1 &,\text{if } j+i \leq m+1\\
          0 &,\text{else},
         \end{cases}
\end{equation}
and $\alpha_{ij}$ representing a symmetric $2\times 2$ matrix $A=(a_{ij})$ located in the lower right corner of $B$.
For example, for $m=5$ we assume that $B$ has the form
\begin{equation}\label{eq:B_a}
B=
\begin{pmatrix}
 1 & 1 & 1 & 1 & 1 \\
 1 & 1 & 1 & 1 & 0 \\
 1 & 1 & 1 & 0 & 0 \\
 1 & 1 & 0 & a_{11} & a_{12}\\
 1 & 0 & 0 & a_{12} & a_{22}
\end{pmatrix},
\end{equation}
and it remains to search through the $2^3=8$ possible values for the $a_{ij}$.
As it is shown in the next section, we find indeed solutions for $B$ which are of this form for values of $m$ up to $24$ (with the only exception that for $m=12,20$ and $21$ we had to take a $3\times 3$ matrix $A$).
It appears that for a $B$ of the form of equation \eqref{eq:B_a}, the global phase $e^{\iE\psi}$ of $U$ determined by equations \eqref{eq:globalbytr} and \eqref{eq:pj} does not depend on the small $A$ matrix, but depends solely on $m$:
\begin{equation}\label{eq:globalphase}
 e^{\iE\psi} = \begin{cases}
               \frac{-1+\iE}{\sqrt{2}} &,\text{for } m \text{ odd}\\
               \iE &,\text{for } m \text{ even},
             \end{cases}
\end{equation}
but we do not have a rigorous proof for general $m$. Note that this implies that the entries of $U$ are roots
of unity of order $4$ for even $m$ and roots of order $8$, but not of order $4$ for odd $m$.

\subsubsection{Equivalence of matrices $B$}
Given some $B$ satisfying conditions i.), ii.) and iii.), it is easy to verify that any matrix $B'= P B P^T$
obtained from $B$ by multiplication with a permutation matrix $P$ also satisfies these conditions.
We are now going to prove the following lemma.
\begin{lem}[Equivalent $B$s]\label{lem:equib}\hfill\\
The complete set of cyclic MUBs specified by the matrices $\{ \mathbbm{1}_d, U', U'^2, \dots , U'^d  \}$ with
$U'$ denoting the Clifford unitary corresponding to $C'=\bigl( \begin{smallmatrix} B' & \mathbbm{1}_d\\ \mathbbm{1}_d & 0_d \end{smallmatrix} \bigr)$ and $B'= P B P^T$ for some permutation matrix $P$ is equivalent
to the complete set of cyclic MUBs specified by $\{ \mathbbm{1}_d, U, U^2, \dots , U^d  \}$ with $U$ denoting
the Clifford unitary corresponding to $C=\bigl( \begin{smallmatrix} B & \mathbbm{1}_d\\ \mathbbm{1}_d & 0_d
\end{smallmatrix} \bigr)$.
\end{lem}
\begin{proof}
From
\begin{equation}
C' = \begin{pmatrix} P & 0_d\\ 0_d & P \end{pmatrix} C \begin{pmatrix} P^T & 0_d\\ 0_d & P^T \end{pmatrix}
\end{equation}
we obtain the corresponding equation $U' = V U V^{-1}$ with the $d=2^m$ dimensional Clifford group unitary $V$ corresponding to the symplectic matrix $\bigl( \begin{smallmatrix} P & 0_d\\ 0_d & P \end{smallmatrix} \bigr)$.
Since the stabilizer state $\ket{\vec{0}}_L$ of the latter matrix is given by $\ket{\vec{0}}\equiv\ket{00\dots0}$, it follows that $V$ is also a permutation matrix.
Hence, according to definition \ref{def:uequi},
\begin{equation}
 U'^k = V U^{\pi(k)} W_k
\end{equation}
for all $1\leq k \leq d+1$, with $\pi(k)=k$ and $W_k=V^{-1}$.
\end{proof}
Since there are $m!$ possible permutation matrices, it is obvious that we may get up to $m!$ equivalent
cyclic MUBs with this method.

\section{Results}\label{sec:results}

Performing the search for a symplectic matrix $C \in M_{2m}(\mathbb{F}_2)$ defining a Clifford unitary $U$ of dimension $d=2^m$ satisfying the conditions of subsection \ref{subsec:constr}, we obtained such $C$'s for $m=1,2,\dots,24$.
Each of these $C$'s is of the form of equation \eqref{eq:CgleichB110} and defines a complete set of cyclic MUBs via the corresponding unitary $U$ given by
equations \eqref{eq:UHdiap} and \eqref{eq:globalphase}.
As a reminder, the search for $C$ reduces to a search for a matrix $B \in M_m(\mathbb{F}_2)$ satisfying the three conditions of subsection \ref{subsec:findc}:
\begin{enumerate}[i.)]
 \item $B=B^T$,
 \item $f_{j}(B)$ is invertible for all $1\leq j \leq 2^{m-1}$,
 \item $f_{2^{m-1}}(B)=f_{2^{m-1}-1}(B)$.
\end{enumerate}

\subsection{The case $m=1$}
The matrix $B$ is a scalar now and $f_1(B)=B$ has to be equal to $f_0(B)=1$ which leads to the single solution $B=1$ and
$C = \bigl(\begin{smallmatrix} 1 & 1 \\ 1 & 0 \end{smallmatrix}\bigr)$.
The corresponding unitary $U$ is given by
\begin{equation}
 U = H \cdot  \operatorname{diag}( 1 ,-\iE ) \cdot \frac{-1+\iE}{\sqrt{2}}  =
 \frac{-1+\iE}{2}  \begin{pmatrix} +1 & -\iE \\ +1 & +\iE \end{pmatrix}
\end{equation}
and has the eigenvalues $\{ \omega, \omega^2 \}$ with $\omega = \exp\bigl( 2\pi \iE / 3 \bigr)$.
Note that $U$ can also be expressed as 
\begin{equation}
 U=-\exp\bigl(-\iE\pi(X+Y+Z)/(3 \sqrt{3})\bigr),
\end{equation}
which is a rotation on the
Bloch sphere around the axis $(1,1,1)$ with rotation angle $2\pi/3$ that corresponds to the operator $T$ used by Gottesman and Lo \cite{G98, Lo01,GL03}.

\subsection{The case $m=2$}
The $2\times 2$ matrix $B$ has to fulfill the condition that $f_2(B)=\mathbbm{1}_2+B^2$ equals $f_1(B)=B$ which leads to the equation $\mathbbm{1}_2+B+B^2=0_2$ having $m!=2$ symmetric solutions, which can be obtained via $B'=PBP^T$ from the matrix
\begin{equation}
B = \begin{pmatrix} 1 & 1 \\ 1 & 0 \end{pmatrix},
\end{equation}
by applying all $m!$ permutation matrices $P$.
The unitary $U$ corresponding to the matrix $C = \bigl(\begin{smallmatrix} B & \mathbbm{1}_2 \\ \mathbbm{1}_2 & 0_2 \end{smallmatrix}\bigr)$ is given by
\begin{equation}
\begin{split}
 U &= H^{\otimes 2} \cdot  \operatorname{diag}( 1,1,-\iE,\iE ) \cdot \iE  \\
   &=\frac{1}{2}\begin{pmatrix}
      +\iE &+\iE &+1 & -1 \\
      +\iE &-\iE &+1 & +1 \\
      +\iE &+\iE &-1 & +1 \\
      +\iE &-\iE &-1 & -1
     \end{pmatrix},
\end{split}
\end{equation}
and has the eigenvalues $\{ \omega, \omega^2, \omega^3, \omega^4 \}$ with $\omega = \exp\bigl( 2\pi \iE / 5 \bigr)$.

\subsection{The case $m=3$}
The $3\times 3$ matrix $B$ has to fulfill the condition that $f_4(B)=\mathbbm{1}_3+B^2+B^4$ is equal to $f_3(B)=B^3$ which leads to
$\mathbbm{1}_3+B^2+B^3+B^4 = 0_3$, or using factorization modulo $2$,
$(\mathbbm{1}_3+B+B^3) \cdot (\mathbbm{1}_3+B) = 0_3$.
Since in addition $f_{2}(B)=\mathbbm{1}_3+B^2 = (\mathbbm{1}_3+B)^2$ has to be invertible, $(\mathbbm{1}_3+B)$ has to be invertible as well.
Hence, all valid matrices $B$ satisfy
\begin{equation}
 \mathbbm{1}_3+B+B^3=0_3  \text{ and } B=B^T.
\end{equation}
A computer search reveals that there are $m!=6$ such matrices, which can be obtained from
\begin{equation}
 B = \begin{pmatrix} 1 & 1 & 1 \\ 1 & 1 & 0 \\ 1 & 0 &0  \end{pmatrix},
\end{equation}
via $B'=PBP^T$ by applying all $m!$ permutation matrices $P$.
The unitary $U$ corresponding to $C = \bigl(\begin{smallmatrix} B & \mathbbm{1}_3 \\ \mathbbm{1}_3 & 0_3 \end{smallmatrix}\bigr)$
is given by
\begin{equation}
\begin{split}
 U &= H^{\otimes 3} \cdot  \operatorname{diag}( 1,1,-\iE,-\iE,-\iE,\iE,1,-1 ) \cdot \frac{-1+\iE}{\sqrt{2}} \\
   &=\frac{-1+\iE}{4}\begin{pmatrix}
      +1 &+1 &-\iE &-\iE &-\iE &+\iE &+1 &-1 \\
      +1 &-1 &-\iE &+\iE &-\iE &-\iE &+1 &+1 \\
      +1 &+1 &+\iE &+\iE &-\iE &+\iE &-1 &+1 \\
      +1 &-1 &+\iE &-\iE &-\iE &-\iE &-1 &-1 \\
      +1 &+1 &-\iE &-\iE &+\iE &-\iE &-1 &+1 \\
      +1 &-1 &-\iE &+\iE &+\iE &+\iE &-1 &-1 \\
      +1 &+1 &+\iE &+\iE &+\iE &-\iE &+1 &-1 \\
      +1 &-1 &+\iE &-\iE &+\iE &+\iE &+1 &+1 
     \end{pmatrix}.
\end{split}
\end{equation}
The eigenvalues of $U$ are given by $\{ \omega^k \vert k=1,2,\dots,8 \}$ with $\omega = \exp\bigl( 2\pi \iE / 9 \bigr)$.
According to lemma \ref{lem:equib}, the Clifford unitaries $U$ corresponding to the remaining five matrices $B$ generate equivalent sets of MUBs.

\subsection{The case $m=4$}
The $4\times 4$ matrix $B$ has to fulfill the condition that $f_8(B)=\mathbbm{1}_4 +B^4 +B^6 +B^8$ equals $f_7(B)=B^7$.
Applying factorization modulo $2$, we find that all valid matrices $B$ satisfy $B=B^T$ and
\begin{equation}
 (\mathbbm{1}_4 + B + B^4)\cdot (\mathbbm{1}_4 + B + B^2 + B^3 + B^4) = 0_4.
\end{equation}
It turns out that there are $96$ such matrices $B$,
which can be grouped into two sets of $48$ matrices each:

The solutions of the first set satisfy $\mathbbm{1}_4 + B + B^4=0_4$ and $B^{15}=\mathbbm{1}_4$ and can be further divided into two subsets, one of which is given by the matrices $B'=PBP^T$ with $P$ denoting the $m!=24$ permutation matrices and
\begin{equation}
B = \begin{pmatrix} 
  1 & 1 & 1 & 1\\
  1 & 1 & 1 & 0\\
  1 & 1 & 0 & 1\\
  1 & 0 & 1 & 0
\end{pmatrix},
\end{equation}
and the other one given by the matrices $B'=PBP^T$ with
\begin{equation}
B = \begin{pmatrix} 
  1 & 1 & 1 & 0\\
  1 & 0 & 0 & 0\\
  1 & 0 & 0 & 1\\
  0 & 0 & 1 & 1
\end{pmatrix}.
\end{equation}

The solutions of the second set satisfy $\mathbbm{1}_4 + B + B^2 + B^3 + B^4 = 0_4$ and $B^5=\mathbbm{1}_4$ and can be further divided into two subsets, one of which is given by the $m!=24$ matrices $B'=PBP^T$ and
\begin{equation}\label{eq:b4}
B = \begin{pmatrix} 
  1 & 1 & 1 & 1\\
  1 & 1 & 1 & 0\\
  1 & 1 & 0 & 0\\
  1 & 0 & 0 & 1
\end{pmatrix},
\end{equation}
and the other one given by the $m!=24$ matrices $B'=PBP^T$ with
\begin{equation}
B = \begin{pmatrix} 
  1 & 1 & 1 & 0\\
  1 & 0 & 0 & 1\\
  1 & 0 & 0 & 0\\
  0 & 1 & 0 & 0
\end{pmatrix}.
\end{equation}

\subsection{The cases $m \geq 4$}

For $m\geq 4$ we assume that there exists a matrix $B$ of the form of equation \eqref{eq:B_a}, and we check whether one of the eight symmetric $2\times 2$ matrices $A$ leads to a $B$ satisfying conditions i--iii.
If for a particular value of $m$ no such $a$ is found, we increase the dimension of $a$ and search for a suitable $3\times 3$ matrix.
In table \ref{table1} we present suitable matrices $A$ for $m=4,\dots,24$.
\begin{table}
\begin{tabular}{c|ccccccc}
$m$ & $4$ & $5$ & $6$ & $7$ & $8$ & $9$ & $10$ \\
\hline
$A$ &
$\bigl(\begin{smallmatrix}0&0\\0&1\end{smallmatrix}\bigr)$ & 
$\bigl(\begin{smallmatrix}0&0\\0&0\end{smallmatrix}\bigr)$ & 
$\bigl(\begin{smallmatrix}0&0\\0&0\end{smallmatrix}\bigr)$ & 
$\bigl(\begin{smallmatrix}0&0\\0&1\end{smallmatrix}\bigr)$ & 
$\bigl(\begin{smallmatrix}0&1\\1&1\end{smallmatrix}\bigr)$ & 
$\bigl(\begin{smallmatrix}0&0\\0&0\end{smallmatrix}\bigr)$ & 
$\bigl(\begin{smallmatrix}1&0\\0&0\end{smallmatrix}\bigr)$   
\end{tabular}
\begin{tabular}{c|ccccccc}
$m$ & $11$ & $12$ & $13$ & $14$ & $15$ & $16$ & $17$ \\
\hline
$A$ &
$\bigl(\begin{smallmatrix}0&0\\0&0\end{smallmatrix}\bigr)$ & 
$\Bigl(\begin{smallmatrix}0&0&1\\0&0&0\\1&0&0\end{smallmatrix}\Bigr)$ & 
$\bigl(\begin{smallmatrix}0&0\\0&1\end{smallmatrix}\bigr)$ & 
$\bigl(\begin{smallmatrix}0&0\\0&0\end{smallmatrix}\bigr)$ & 
$\bigl(\begin{smallmatrix}0&1\\1&1\end{smallmatrix}\bigr)$ & 
$\bigl(\begin{smallmatrix}0&0\\0&1\end{smallmatrix}\bigr)$ & 
$\bigl(\begin{smallmatrix}0&0\\0&1\end{smallmatrix}\bigr)$   
\end{tabular}
\begin{tabular}{c|ccccccc}
$m$ & $18$ & $19$ & $20$ & $21$ & $22$ & $23$ & $24$   \\
\hline
$A$ &
$\bigl(\begin{smallmatrix}0&0\\0&0\end{smallmatrix}\bigr)$ & 
$\bigl(\begin{smallmatrix}0&0\\0&1\end{smallmatrix}\bigr)$ & 
$\Bigl(\begin{smallmatrix}1&0&0\\0&0&0\\0&0&1\end{smallmatrix}\Bigr)$ & 
$\Bigl(\begin{smallmatrix}0&0&1\\0&0&0\\1&0&0\end{smallmatrix}\Bigr)$ & 
$\bigl(\begin{smallmatrix}1&0\\0&0\end{smallmatrix}\bigr)$ & 
$\bigl(\begin{smallmatrix}0&0\\0&0\end{smallmatrix}\bigr)$ & 
$\bigl(\begin{smallmatrix}1&0\\0&1\end{smallmatrix}\bigr)$   
\end{tabular}
\caption{This table shows a matrix $A$ which corresponds to the lower right corner of a matrix $B\in M_m(\mathbb{F}_2)$ of the form of equation \eqref{eq:B_a} satisfying conditions i--iii.\label{table1}}
\end{table}
According to this table for $m=4$ for example, a suitable $B$ is given by \eqref{eq:b4}
and the unitary $U$ corresponding to $C = \bigl(\begin{smallmatrix} B & \mathbbm{1}_4 \\ \mathbbm{1}_4 & 0_4 \end{smallmatrix}\bigr)$ is given by
\begin{multline}
 U = H^{\otimes 4} \cdot \operatorname{diag}(
1, -\iE, 1, -\iE, -\iE, -1, \iE, 1, \\
-\iE, 1, \iE, -1, 1, \iE, 1, \iE ) \cdot \iE,
\end{multline}
where we obtained the phases $p_{\vec{j}}$ from $B$ with the help of equation \eqref{eq:pj}.
For values of $m\gtrsim 24$ the test whether condition ii.) is satisfied for a particular matrix $B$ starts to consume a considerable amount of time, preventing us from finding suitable matrices $B$ for values of $m$ higher than $24$.

\section{Conclusion}\label{sec:conclu}
In this paper, we presented a method to construct complete sets of cyclic mutually unbiased bases in
even prime-power dimensions. We used this method to explicitly compute unitaries which generate
such MUBs in all dimensions $2^m$ with $m \in \MgE{24}$, and this limit arises only due to limits of
computational power. We have reason to believe that is is possible to prove the existence of at least one
suitable matrix $B$ as in sections \ref{sec:constr} and \ref{sec:results} for every $m \in \N$, which would
yield a simple proof for the existence of cyclic MUBs in these dimensions, but this is not within the
scope of this work.

\appendix

\section{Properties of the polynomials $f_k$}\label{sec:propfn}
In this appendix, we prove some properties of the polynomials $f_k$ over the finite field $\mathbb{F}_2$,
defined by $f_{-2}(x) = 1$, $f_{-1}(x) = 0$ and $f_k(x) = f_{k-1}(x)\cdot x + f_{k-2}(x)$ for $k \in \N_0$.
By this recursion, it is obvious that $f_k$ is a normalized polynomial of degree $k$ for $k \in \N_0$. Its coefficients are determined in the following lemma.
\begin{lem}[Coefficients of the polynomial $f_k$]\label{KoeffPol}\hfill\\
  For $k \in \N_0$ there holds $f_k(x) = \sum_{i = 0}^{k} a_i^{(k)} x^i$ with
  \begin{equation*}
    a_i^{(k)} = \begin{cases}
                  \binom{(k+i)/2}{(k-i)/2} \mod 2, &\text{ if $i \equiv k \mod 2$},\\
                  0,                               &\text{ otherwise}.
                \end{cases}
  \end{equation*}
  In other words, $f_k(x) = \sum_{r = 0}^{[k/2]} \left\{\binom{k-r}{r} \mod 2\right\} \cdot x^{k-2r}$.
\end{lem}
\begin{proof}
We have $f_0(x) = 1$ and $f_1(x) = x$, so that the statement holds in these cases.
The recursion formula can be restated as $a_i^{(k)} = a_{i-1}^{(k-1)} + a_i^{(k-2)}$.
In case $i \equiv k \mod 2$, we have to show
\begin{equation*}
  \binom{\frac{k+i}{2}}{\frac{k-i}{2}}
  \equiv \binom{\frac{k+i}{2} - 1}{\frac{k-i}{2}} + \binom{\frac{k+i}{2} - 1 }{\frac{k-i}{2} - 1} \mod 2;
\end{equation*}
but this is a standard result from combinatorics and the case $i \not\equiv k \mod 2$ holds
in a similar fashion.
\end{proof}
We now want to find a criterion, when there holds $\binom{n}{k} = \frac{n!}{(n-k)!\,k!} \equiv 0 \mod 2$
for a  binomial coefficient. For this, let $\mathbb{P}$ be the set of prime numbers and denote by
$[x] := \max\Mge{n \in \Z}{n \leq x}$ the Gauss' floor function.
\begin{lem}[Factorization of binomial coefficients]\label{Factorisation}\hfill\\
  For any $n \in \N$ there holds $n! = \prod_{p \in \mathbb{P}} p^{e(n,p)}$ with
  $e(n,p) := \sum_{j = 1}^{\infty} [n/p^j]$; given some $k \in \MgN{n}$ there holds
  $\binom{n}{k} = \prod_{p \in \mathbb{P}} p^{e^\prime(n,k,p)}$ with
  \begin{equation*}
    e^\prime(n,k,p) := \sum\nolimits_{j = 1}^{\infty} \left\{[n/p^j] - [(n-k)/p^j] - [k/p^j]\right\}.
  \end{equation*}
\end{lem}
\begin{proof}
  There are $[n/p]$ multiples of $p$ contained in $n!$ (counted once), $[n/p^2]$ multiples of $p^2$
  (counted twice) etc., which shows the first part. The second part is an immediate consequence thereof.
\end{proof}
Since the terms in the curly bracket are either $0$ or $1$ for any $j$, we have $\binom{n}{k} \equiv 0 \mod 2$,
if and only if at least one term $[n/2^j] - [(n-k)/2^j] - [k/2^j]$ is positive. This is the case, if and only
if $(n-k) \mod 2^j + k \mod 2^j \geq 2^j$ holds for at least one $j \in \N$. We use this fact in the
proof of the next two lemmata.
\begin{lem}[Polynomials $f_{2^m-1}$]\label{lem:f2m1}\hfill\\
  If $k = 2^m - 1$ for some $m \in \N$, then $f_k(x) = x^k$.
\end{lem}
We shall give a direct proof here; another proof may be obtained from lemma \ref{fk-odd}.
\begin{proof}
By construction $k$ is odd, and by lemma \ref{KoeffPol} all coefficients $a_i^{(k)} = 0$ with even
$i$ vanish. We have $a_k^{(k)} = 1$ and for the remaining odd $i$, we have to show
$a_i^{(k)} = \binom{\frac{k+i}{2}}{\frac{k-i}{2}} \mod 2 = 0$.
Let us now write $\frac{k-i}{2} = 2^e \cdot r$ for odd $r$ and $e \in \MgE{m-2}$.
We find $\frac{k+i}{2} - \frac{k-i}{2} = i = 2^m -1 - 2^{e+1} \cdot r \equiv -1 \mod 2^{e+1}$, thus
$i \mod 2^{e+1} = 2^{e+1} - 1$ is the maximally possible value and $\frac{k-i}{2} \mod 2^{e+1} \neq 0$.
Therefore, $i \mod 2^{e+1} + \frac{k-i}{2} \mod 2^{e+1} \geq 2^{e+1}$.
\end{proof}
\begin{lem}[Polynomials $f_{2^m-2}$]\label{lem:f2m2}\hfill\\
  For $m \in \N$, there holds $f_{2^m-2}(x) = \sum_{j = 1}^{m}  x^{2^m - 2^j}$.
\end{lem}
\begin{proof}
  Let $k := 2^m - 2$. According to lemma \ref{KoeffPol} all $a_i^{(k)}$ with odd $i$
  vanish. For the even $i$, we consider $\frac{k-i}{2}$ and $i$ in a similar fashion as in lemma \ref{lem:f2m1}.
  We will write these numbers in $m$-bit binary notation, i.\,e. $k = (1 \dots 10)_2$ and
  $i = (i_{m-1}i_{m-2}\dots i_1 0)_2$; we have $i_0 = 0$, since $i$ is even, and we use commata as appropriate.
  Therefore $\frac{k-i}{2} = (0, 1-i_{m-1}, 1-i_{m-2}, \dots, 1-i_2, 1-i_1)_2$.
  The condition $\frac{k-i}{2} \mod 2^j + i \mod 2^j \geq 2^j$ holds, if at the $j$-th position,
  $j \in \MgE{m-2}$, there occurs an overflow, i.\,e. if
  $(1-i_{j+1}) + i_j > 1$ or $i_{j+1} = 1 \wedge i_j = 0$. This holds for no choice of $j$, if and only
  if $i$ is of the form $(1\dots 10\dots 0)_2$, i.\,e. $i = 2^m - 2^j$ for some $j$.
\end{proof}

\begin{lem}[Block-determinants of polynomials]\label{lem:fjk}\hfill\\
  For the polynomials $f_k$, there holds
  \begin{equation*}
    f_k(x)f_{l-1}(x) - f_l(x)f_{k-1}(x) = f_{\betrag{k-l}-1}(x).
  \end{equation*}
\end{lem}
\begin{proof}
  Let $u(k,l) := f_k(x)f_{l-1}(x) - f_l(x)f_{k-1}(x)$. By the recursion $f_k(x) = f_{k-1}(x)x + f_{k-2}(x)$
  we may write
  \begin{equation*}
    f_k(x)f_{l-1}(x) = f_{k-1}(x) f_{l-1}(x)x + f_{k-2}(x) f_{l-1}(x)
  \end{equation*}
  and similarly for $k$ and $l$ exchanged. Subtracting these terms yields $u(k,l) = -u(k-1,l-1)$.
  Assuming without loss of generality $k \geq l$, this results in $u(k,l) = (-1)^l u(k-l,0)$. Since
  $f_{-1}(x) = 0$ and $f_0(x) = 1$, there holds $u(k,l) = (-1)^{l+1} f_{k-l-1}(x)$, and since we work
  over the field $\F_2$, we ignore the prefactor $(-1)^{l+1}$.
\end{proof}
In the following, we will consider divisibility properties of the polynomials $f_k$. For this, it is
useful to note the generalized recursion $f_{k+l} = f_k f_l + f_{k-1} f_{l-1}$, which
can be directly read off from eq. \eqref{eq:Cn-fk} or proven by induction.
\begin{lem}[Divisibility of polynomials]\hfill\\
  If $k^\prime \in \N$ divides $k \in \N$, then $f_{k'-1}$ divides $f_{k-1}$.
\end{lem}
\begin{proof}
  Let $k = nk^\prime$ for an appropriate $n \in \N$. We note that the case $n = 1$ is trivial and proceed
  by induction \linebreak over $n$. By the generalized recursion, we find that there holds $f_{(n+1)k^\prime-1}
  = f_{nk^\prime + (k^\prime - 1)} = f_{nk^\prime} f_{k^\prime - 1} + f_{nk^\prime - 1} f_{k^\prime - 2}$,
  where $f_{k^\prime - 1}$ and $f_{nk^\prime - 1}$ are divisible by $f_{k^\prime - 1}$.
\end{proof}
\begin{lem}[Factorization of $f_k$ for odd $k$]\label{fk-odd}\hfill\\
  There holds $f_{2k+1}(x) = f_k(x)^2 \cdot x$ for $k \in \N_0$. If we set $k + 1 = 2^e \cdot r$
  for odd $r$, we have $f_k(x) = f_{r-1}(x)^{2^e} \cdot x^{2^e-1}$.
\end{lem}
\begin{proof}
  By the generalized recursion, we find the relation
  $f_{2k+1} = f_{(k+1)+k} = f_{k+1} f_k + f_k f_{k-1} = f_k (f_{k+1} + f_{k-1})$
  and use the fact that $f_{k+1}(x) + f_{k-1}(x) = x f_k(x)$. The second part follows by
  induction over $e$.
\end{proof}
Thus, in order to check for invertibility of $f_k(x)$ for all $k \in \MgN{k_{\max}}$ for some $k_{\max} \in \N$
(as in condition ii.) of the main text), we only have to check invertibility of $x$ itself and the
$f_k(x)$ with even $k$. If we define polynomials $g_k(x) := \sum_{i = 0}^{k} b_i^{(k)} x^i$ with
$b_i^{(k)} := \binom{k+i}{k-i} \mod 2$, we have $f_{2k}(x) = g_k(x^2)$, and we only have to deal with this
reduced set of polynomials in $x^2$.

\section{Eigenvalues of matrices which generate cyclic MUBs}\label{sec:ewcyclicU}
Let us consider a unitary matrix $U \in M_d(\C)$ which generates a complete set of cyclic MUBs.
We were not able to give a formal proof for the following conjecture, but our results indicate that it
may be true for all matrices produced by our method.
\begin{con}[Spectrum of generators of MUBs]\label{SpectrumConjecture}\hfill\\
  Let $U$ be a generator of a complete set of cyclic MUBs. Then, its spectrum is non-degenerate and
  consists of all roots of unity of order $d+1$ with precisely one exception.
\end{con}
By definition, $U^{d+1} = \Eins_d$, i.\,e., all eigenvalues of $U$ are roots of unity of order $d+1$,
and the second part follows immediately from the non-degeneracy. Since we may multiply $U$ with an
arbitrary power of $\omega = \exp\bigl(\frac{2\pi\iE}{d+1}\bigr)$, we may choose $1$ not to lie in the
spectrum. In this case, we have $\tr U = -1$.

\section{Results from algebra}
Let $R$ be a commutative ring with identity and consider a matrix $A \in M_m(R)$.
We then define the \emph{complementary matrix} of $A$ as $\tilde A = (\tilde a_{ij})_{i,j=1}^{m} \in M_m(R)$
with coefficients $\tilde a_{ij} = (-1)^{i+j} \det A_{ji}$, where $A_{ji} \in M_{m-1}(R)$ is
constructed from $A$ by removing the $j$-th row and the $i$-th column. We then have the following
criterion for \mbox{invertibility} of a matrix; cf. e.\,g. Hungerford \cite{Hungerford}, Prop.~VII.3.7 on p. 353,
or Bourbaki \cite{Bourbaki}, \S 8.6, Prop. 12 on p. III.99.
\begin{thm}[Cramer's rule]\label{Cramer}\hfill\\
  Let $R$ be a commutative ring with identity. For every matrix $A \in M_m(R)$, there holds
  $\tilde A A = A \tilde A = (\det A) \Eins_m$. In particular, $A$ is invertible, if and only
  if $\det A$ is an invertible element in $R$.
\end{thm}
\begin{proof}
  The main statement follows by direct calculation, and we have $\det A \cdot \det A^{-1} = \det (AA^{-1})
  = \det \Eins_m = 1$, if $A$ is invertible.
\end{proof}
In this paper, we need only the following lemma.
\begin{lem}[Invertible Matrices]\label{lem:invmatrices}\hfill\\
  Let $A,\,B,\,C,\,D \in M_m(\F_2)$ be commuting matrices. Then the block matrix
  $\bigl(\begin{smallmatrix} A & B \\ C & D \end{smallmatrix}\bigr) \in M_{2m}(\F_2)$ is invertible, if and only
  $AD-BC \in M_m(\F_2)$ is.
\end{lem}
\begin{proof}
Let $R$ be the commutative subring with identity generated by the elements $A,\,B,\,C,\,D \in M_m(\F_2)$
in the matrix ring $M_{2m}(\F_2)$. Then the statement follows immediately from Theorem~\ref{Cramer}.
\end{proof}

\begin{acknowledgments}
The authors thank Chris Charnes for drawing their attention to the problem of constructing complete sets
of cyclic MUBs and Chris Charnes and Gernot Alber for helpful discussions.
Financial support by CASED is acknowledged.
\end{acknowledgments}

\bibliography{references}

\end{document}